\input psfig.sty
\def\onee{1E1724--3045}
\def\nuflat{$\nu_{\rm flat}$}

\def\mzero{M$_{\rm 0}$}
\def\nult{$\nu_{\rm LT}$}
\def\nucl{$\nu_{\rm CL}$}
\def\nus{$\nu_{\rm s}$}
\def\nuk{$\nu_{\rm k}$}
\def\iquacin{I$_{45}$}
\def\rossi{{\it Rossi X-ray Timing Explorer~}}
\def\pca{{\it Proportional Counter Array}}
\def\ergs{ergs s$^{-1}$}
\def\ergscm{ergs s$^{-1}$ cm$^{-2}$}

\def\msol{$\rm M_\odot$}

\def\ctss{cts s$^{-1}$}
\def\nuqpo{$\nu_{\rm QPO}$}
\def\groj{GRO J0422+32}

\documentclass[epsfig]{aa}

\usepackage{astron}

\begin{document}

   \thesaurus{06     % A&A Section 6: Form. struct. and evolut. of stars
              (08.14.1;
               08.02.3 
               10.07.2;
               13.25.3;
               13.25.5;
               13.07.2)} % Stars: structure of.
   \title{RXTE Observation of the X-ray burster 1E1724-3045}

   \subtitle{I. Timing study of the persistent X-ray emission with the PCA}

   \author{J. F. Olive\inst{1}, D. Barret\inst{1}, L. Boirin\inst{1},
           J. E. Grindlay\inst{2}, J. H. Swank\inst{3} and A. P.
Smale\inst{3}
          }

   \offprints{D. Barret}

   \institute{Centre d'Etude Spatiale des Rayonnements, CNRS/UPS, 
9 Avenue du Colonel Roche, 31028 Toulouse Cedex 04, France 
(email: barret@cesr.fr)
         \and
             Harvard Smithsonian Center for Astrophysics, 60 Garden Street,
             Cambridge, MA 02138, USA
          \and 
Laboratory for High Energy Astrophysics, NASA Goddard Space Flight Center
Greenbelt, MD 20771, USA
             }

   \date{Received ; accepted}

\titlerunning{RXTE/PCA observations of 1E1724-3045}
\authorrunning{Olive et al.}

  \maketitle

\begin{abstract} 

We report on the \rossi~Proportional Counter Array (PCA) observation
of the X-ray burster \onee~located in the globular cluster Terzan 2.
The observation lasted for about 100 kiloseconds and spanned from
November 4th to 8th, 1996.  The PCA source count rate in the 2-20 keV
range was about 470 cts/s. No large spectral variations were observed
within our observation as inferred from color-color and
hardness-intensity diagram analysis.  The persistent X-ray emission
shows a high level of noise variability, the so-called High Frequency
Noise (HFN) with a fractional Root Mean Squared (RMS) of $\sim$ 25\%
in the $2 \times 10^{-3} - 40$ Hz range.  The strong HFN together with
the hardness of its X-ray spectrum suggest that \onee~is an ``Atoll''
source which was in its ``Island'' state during the observation.  The
Fourier Power Density Spectra (PDS) can be modeled in terms of the sum
of two ``shot noise'' components for which the shots have a
single-side exponential shape (i.e. instantaneous rise and exponential
decay). The characteristic shot decay timescales inferred from the
best fitting of the PDS are $\sim 680$ and 16 msec respectively. The
two components contribute similarly to the total RMS ($\sim$
15\%). The PDS contains also a third component: a broad and asymmetric
peaked noise feature centered at 0.8 Hz.  This Quasi Periodic
Oscillation-like (QPO) feature contributes at the level of $\sim$ 10\%
to the total RMS. Neither the shot timescales nor the QPO frequency
vary with energy. On the other hand, the integrated RMS of all three
components shows a positive correlation with energy up to at least 40
keV. In addition, in the 2--20 keV energy band where the signal to
noise ratio is the highest, we found evidence for a high frequency
component which shows up in the PDS above 100 Hz. In terms of the shot
noise model, a Lorentzian fit of this last component implies a shot
decay timescale of $\sim 1.4$ ms.

We also show that \onee~ has striking timing similarities with the
black hole candidate GRO J0422+32 (Nova Persei 1992).  Our observation
demonstrates that a low frequency QPO simultaneously with a high level
of RMS is not a timing signature unique to black holes.  This extends
the growing list of similarities between Atoll sources and black hole
systems.

\end{abstract}

%
%________________________________________________________________

\section{Introduction}

An X-ray burst from a region including the globular cluster Terzan 2
was first observed by OSO-8 \cite{swank77apjl}. A weak persistent
X-ray source was also found by OSO-8, and Grindlay
\cite*{grindlay78apjl} showed that both the position of this source
and a revised position of the {\it Uhuru} source 4U1722-30
\cite{forman78apjs} were consistent with that of the cluster.  Later
Grindlay et al. \cite*{grind80apjl} using the EINSTEIN HRI instrument
positioned the X-ray burst source inside the core of Terzan 2. Its
name then became 1E1724--3045. Basically reliable spectral
observations (EXOSAT and TTM) show that the source has a rather hard
power law spectrum (photon index around 2) in the 1-20 keV range with
an X-ray flux ranging from 3 to $9 \times 10^{-10}$ \ergscm.  There is
also an indication from EINSTEIN data that at higher flux level ($\sim
2.1 \times 10^{-9}$ \ergscm, i.e. in its high state), the X-ray
spectrum softens and is better described by a Bremsstrahlung model
with a temperature of $\sim 6$ keV
\cite{tz2:barret98aa}.

%The X-ray burster 1E1724-3045 located in the globular cluster Terzan
%2, in the direction of the Galactic center, has been repeatedly
%observed by various experiments since its discovery in 1977 by UHURU
%\cite{swank77apjl}.  

Estimates of the cluster distance range from 5.2 to 7.7 kpc
\cite{ortolani97aa}. These values are consistent with the one derived
from a type I X-ray burst that showed photospheric expansion
\cite{grind80apjl,tanaka81}. Adopting 7 kpc, the 1-20 keV low state
luminosity of the source lies around $\sim 4 \times 10^{36}$
\ergs~indicating that \onee~ belongs to the class of low luminosity
systems.

As for the timing properties of the persistent X-ray emission of the
source, very little is known. However, during an EXOSAT observation
performed with the medium energy (ME) experiment, a QPO at 0.092 $\pm
0.001$ Hz was found (pulsed fraction $4.8 \pm 1.2$ \%, Full Width Half
Maximum of $0.053 \pm 0.002$ Hz).  In addition low-frequency noise and the second
harmonic were observed \cite{belli86iau}.  Note however that the
presence of the QPOs reported in the {\it IAU} circular was not
confirmed in a proceedings paper by the same authors
\cite{belli85esa}.

Besides its X-ray properties, \onee~is remarkable by the fact that it
is a source of persistent, though variable, hard X-ray emission ($\sim
35-150$ keV) as observed with SIGMA
\cite{tz2:barret91apjl,gold93:2ndgro}.  \onee~is in fact the first,
and one of the only X-ray bursters (i.e.  weakly magnetized neutron
star) showing such a persistent emission with a spectrum extending up
to $\sim 100$ keV \cite{vargas96int,tavani974thcgro}.

\begin{figure}[t]
\vspace{0cm}
\vspace{0cm}
\hspace{0cm}\centerline{\psfig{figure=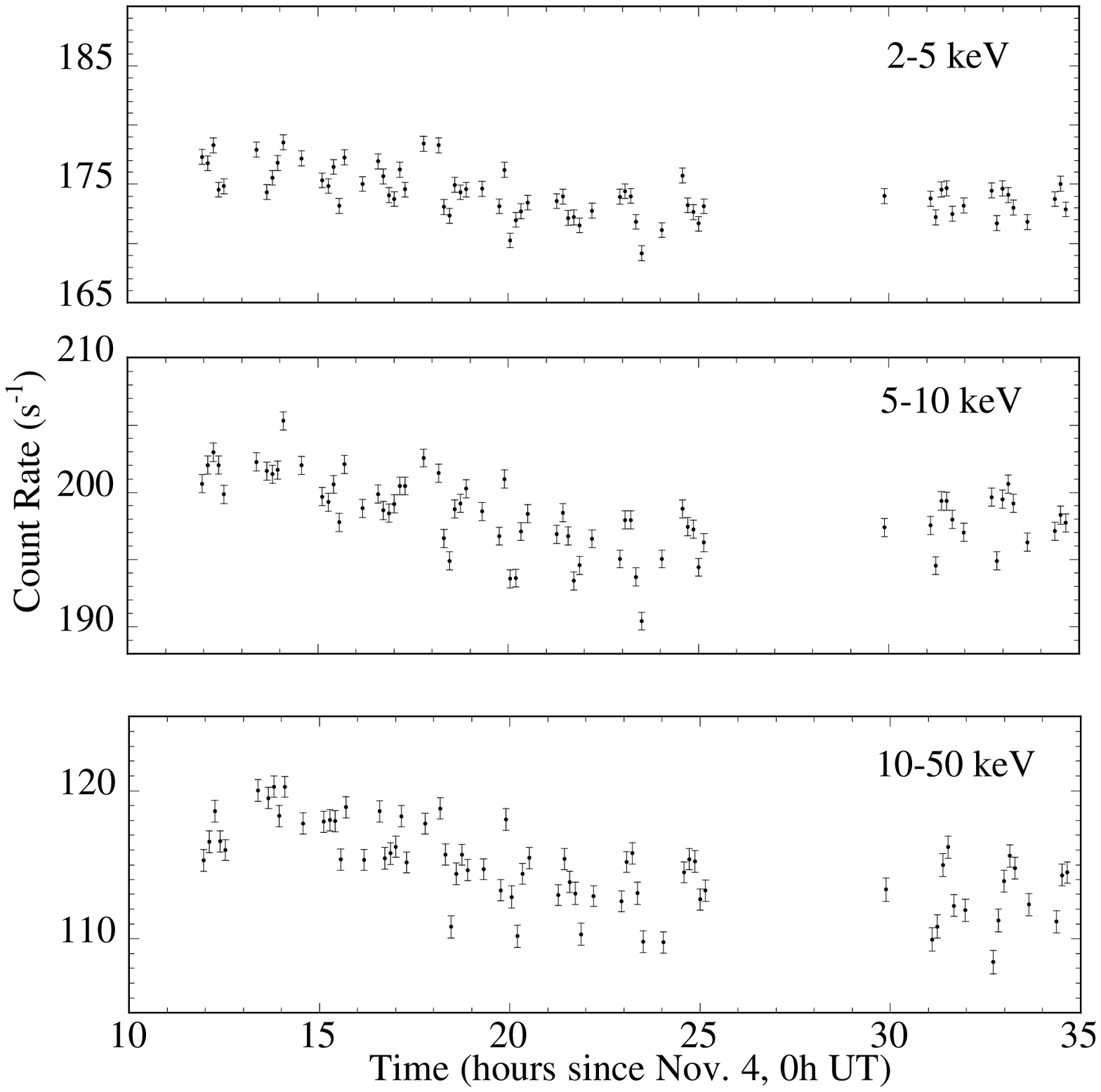,width=9.0cm,height=12cm}}
\vspace{0cm}

\vspace{-4cm}
\hspace{0cm}\centerline{\psfig{figure=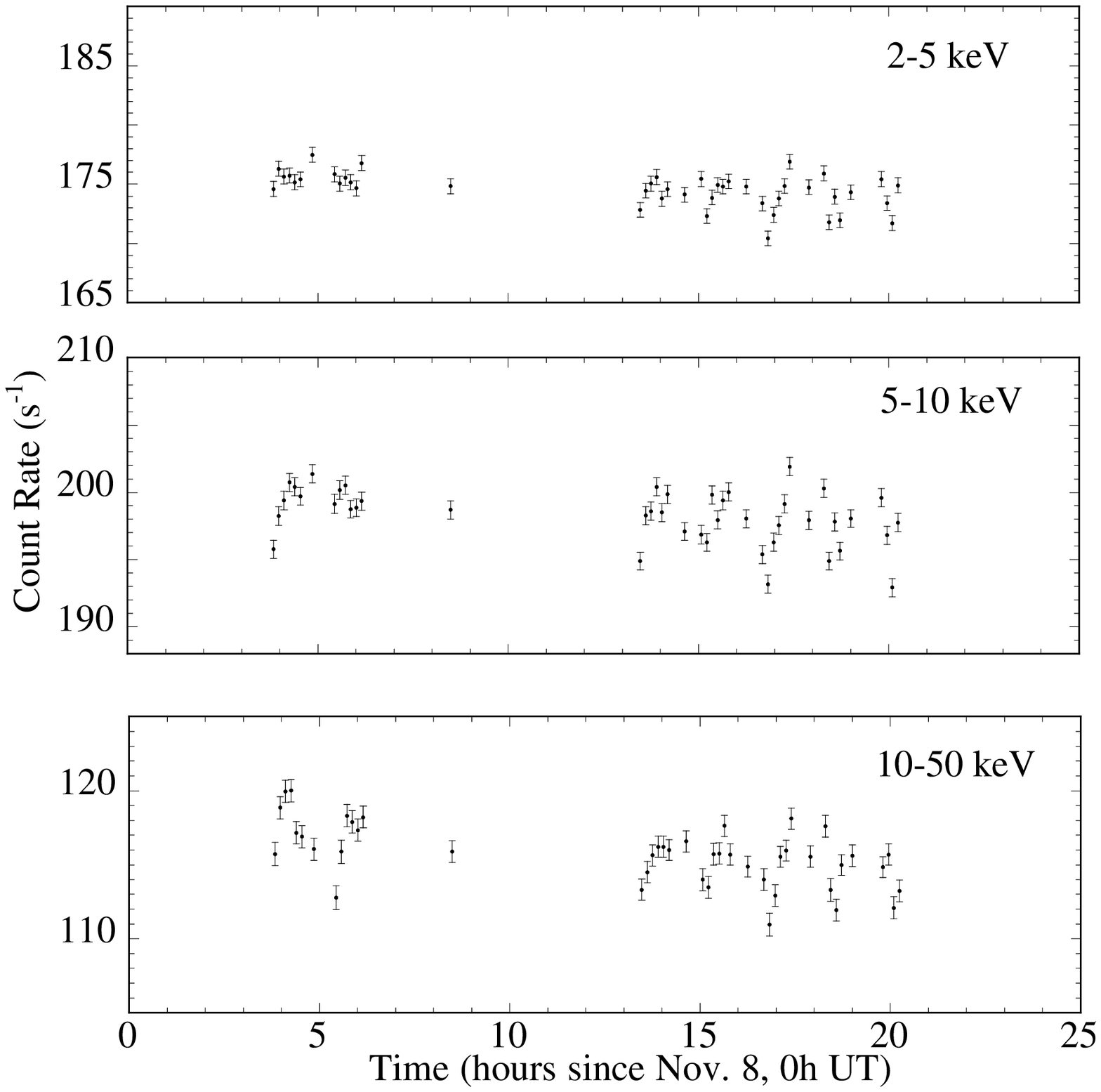,width=9.0cm,height=12cm}}
\vspace{-4cm}

\caption{PCA Light curve of 1E1724-3045 in the energy bands 2-5, 5-10, 
10-50 keV. The bin time is 512 seconds.}

\label{pcalc}
\end{figure}

We proposed \onee~as a target for a Rossi X-ray Timing Explorer
observation for two main reasons: First we wanted to characterize the
rapid variability of the source in order to check whether, as
expected, \onee~is an Atoll source. Second we wished to detail the
continuum energy spectrum simultaneously from X-rays to hard X-rays.

The main focus of this paper is the study of the rapid variability of
the source in the frequency range $2 \times 10^{-3}$ to a few tens of
Hz. The search for high frequency QPOs (above $\ga 300$ Hz) will
appear elsewhere. Similarly the spectral analysis of the combined PCA
and HEXTE data is the subject of a forthcoming paper
\cite{tz2:barret98aa}. The present paper is organized as follows: In
section 2, we present the correlated temporal and spectral analysis
using light curves, color-color and hardness-intensity diagrams.  In
section 3, we present the results of the rapid variability study,
including the detailed modeling of the PDS. In section 4, we discuss
on the possible origins of the fast timing variability of the source,
emphasize the similarities between black hole and neutron star
systems, and address the issue about the nature of the low frequency
QPO observed.

\section{Light curves, color-color and hardness intensity diagrams}

\subsection{The PCA observation}

The RXTE observation took place between November, 4th and 8th, 1996
for a total exposure time of about 100 kiloseconds. The good data
filtered for elevation greater than 10 degrees (as recommended) can be
split in two parts; the first covers from November 4th at 11:54:08
(UT) to November 5th at 10:51:44 (UT) and the second from November 8th
at 00:30:24 (UT) to 23:35:44 (UT).

The \pca~consists of 5 nearly identical large area proportional
counter units (PCU 0 to 4) corresponding to a total collecting area of
$\sim 6500$ cm$^2$ \cite{jahoda96spie}. For safety reasons, PCUs are
switched on and off in the course of an observation; mostly PCUs 3 and
4. As all the PCUs have different energy responses, the absence of a
single one can simulate spectral variations in color-color and
hardness-intensity diagrams. Therefore to avoid this, in this section
we consider only data recorded when the five PCU units are all
working; this corresponds to a net exposure time of $\sim 69$
kiloseconds (elevation greater than 10$^\circ$). There is also an
X-ray burst in the middle of the observation; the analysis of which
will be reported elsewhere. Data recorded around this burst are not
included here (3 orbits of good data are not considered in the present
analysis).

We use the {\it standard 2} data to build light curves, color-color and
hardness-intensity diagrams. For the rapid variability, we use the 16
$\mu$s {\it Science Event} data provided in 64 energy channels
covering the entire PCA energy range. Significant flux from the source
is detected up to $\sim 50$ keV (see Fig. \ref{pcasp}).

\subsection{Light curves, color-color and hardness-intensity diagrams}

We have first generated background subtracted light curves for the
entire observation taking the latest versions of the background models
provided at the RXTE Guest Observer Facilities (GOF) \footnote{We use
the release 1.5 of {\it pcabackest}.}. The light curves of \onee~ in
the energy bands 2--5, 5--10 and 10--50 keV are shown in
Fig. \ref{pcalc} for a binning time of 512 seconds.

As can be seen no large temporal variations were observed during the
observation. However during the first part, its intensity slowly
decreased in all three energy bands. The mean background subtracted
count rate is $\sim 500$ \ctss~ which corresponds to about 40 mCrab in
the 2-50 keV band\footnote{We assume that the Crab produces 2600
\ctss~per PCU \cite{jahoda96spie}}.

Following Hasinger and Van der Klis \cite*{hasinger89aa} we have
generated color-color diagrams using three energy bands: 2-5, 5-10,
and 10-50 keV.  The color-color diagram computed for a binning time of
128 seconds is shown in Fig. 2. The position of the source in the
diagram indicates a rather hard spectrum. The clustering of points in
the diagram indicates that the source spectrum did not change
significantly during the observation.

\begin{figure}
\vspace{0cm}
\hspace{0.5cm}\centerline{\psfig{figure=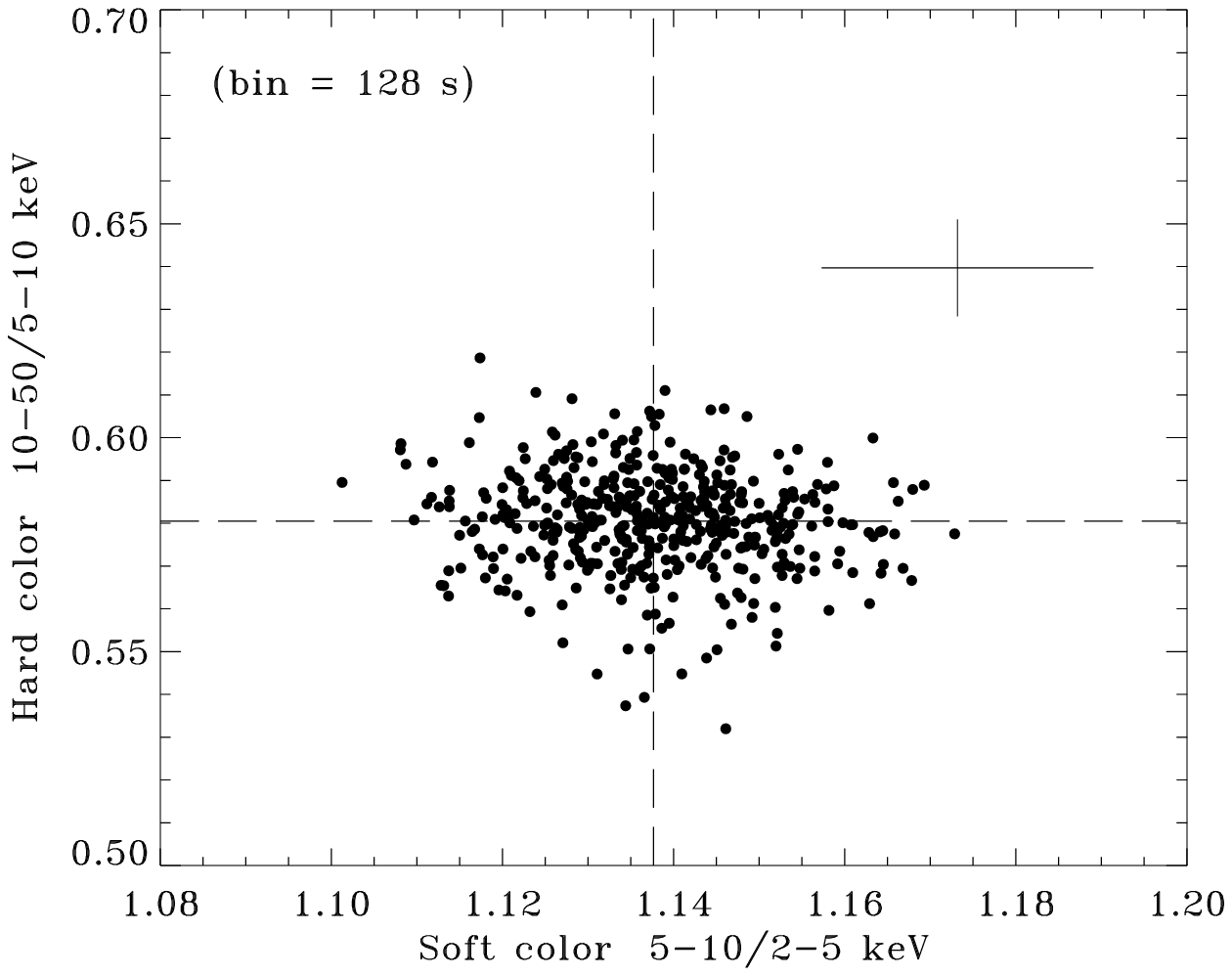,width=11cm}}
\vspace{-1cm}
\label{coco}
\caption{Color-color diagram computed for a binning time of 128 seconds. 
A typical error bar is also shown.}
\end{figure}

\begin{figure}[t!]
\vspace{0cm}
\hspace{0.5cm}\centerline{\psfig{figure=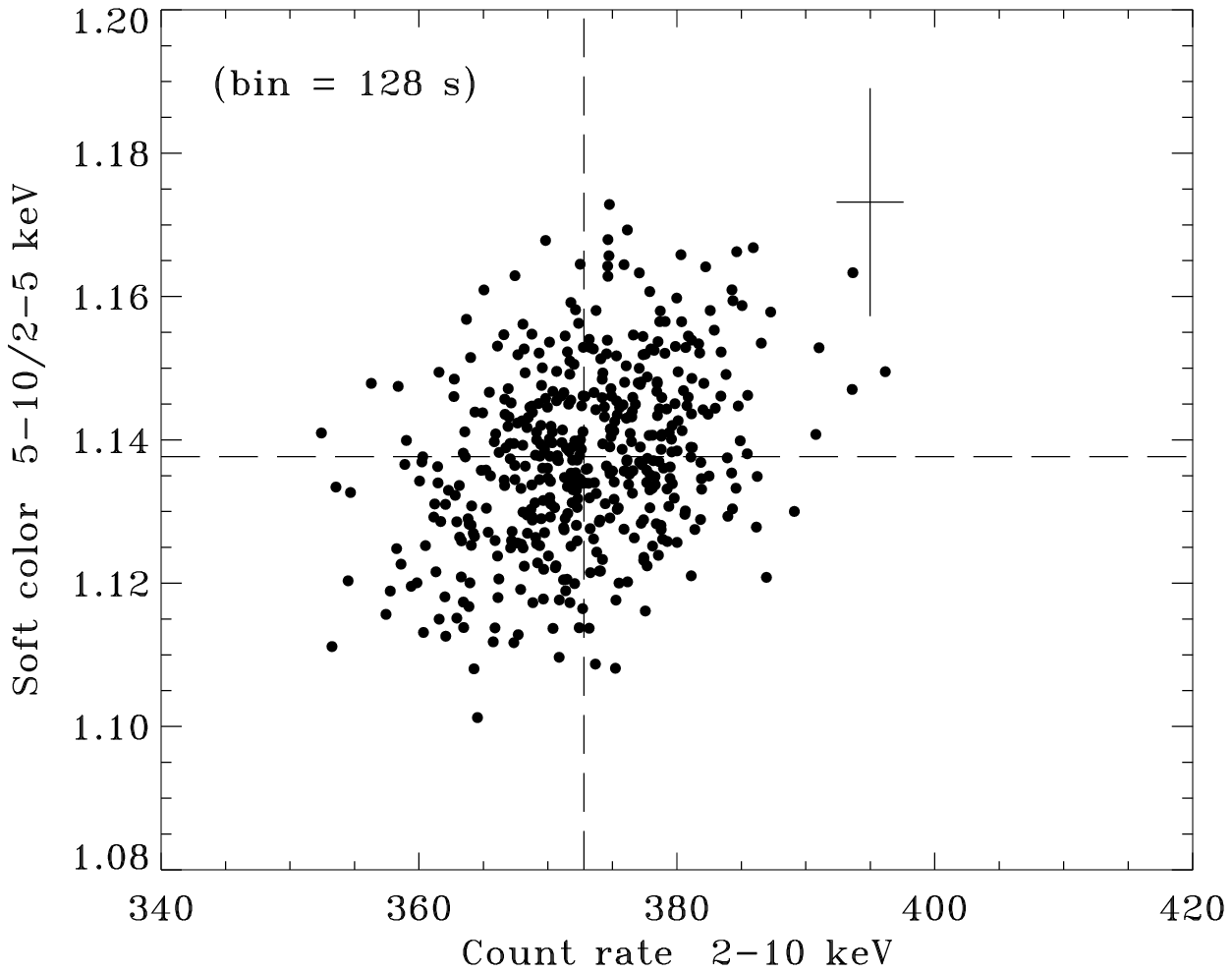,width=11cm}}
\vspace{0cm}

\hspace{0.5cm}\centerline{\psfig{figure=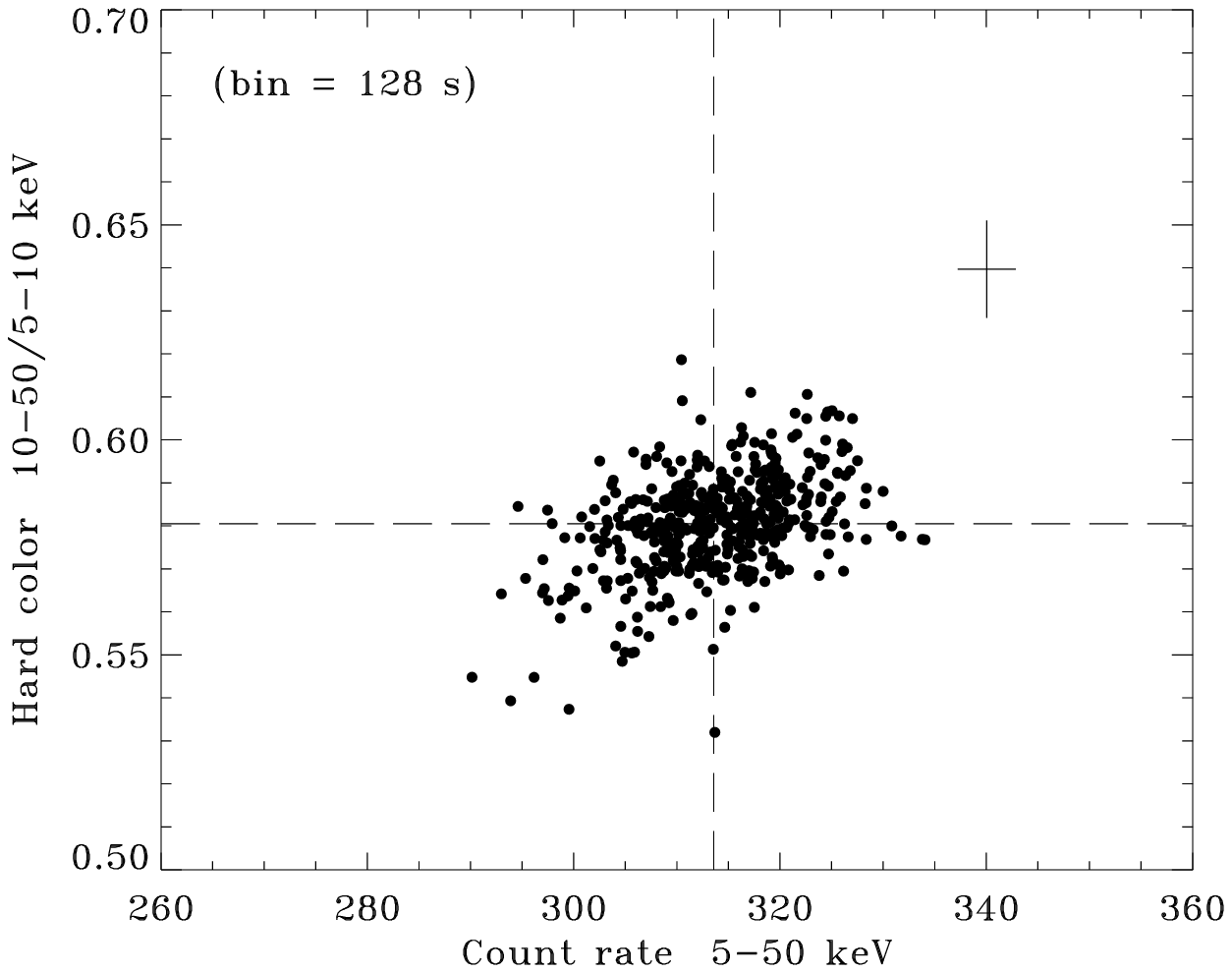,width=11cm}}
\vspace{-1.cm}

\caption{Hardness intensity diagrams (soft and hard) computed for a binning 
time of 128 seconds. A typical error bar is also shown.}
\label{hardnessint}
\end{figure}

Although the source count rate does not differ by more than 10\% along
the observation, we have computed two hardness-intensity diagrams,
again for a binning time of 128 seconds (see Fig
\ref{hardnessint}). There is a weak correlation between both the soft
and hard colours and the count rate. The source moves erratically
within the defined regions. Given the spectral resolution of the PCA,
these variations are hardly detectable with standard spectral analysis
of the continuum spectrum.

\subsection{PCA Energy spectrum}

\begin{figure}[t!]
\vspace{0cm}
\hspace{0.cm}\centerline{\psfig{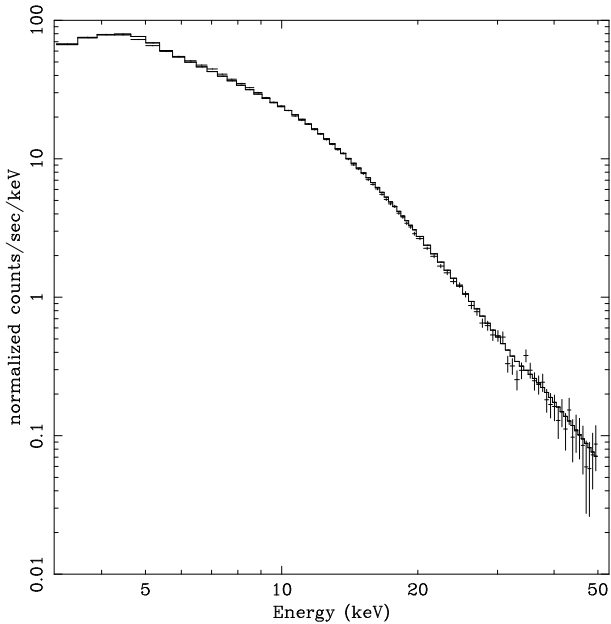}}
\vspace{0cm}
\caption{PCA count spectrum and folded model showing that the source 
is clearly detected up to about 50 keV. Within the calibration
uncertainties at low energies, the spectrum is well fitted with a
power law model of photon index $\sim 2.0$. The fit was performed in
the 3-50 keV band with 2\% systematic errors included in the count
spectrum (reduced $\chi^2=1.29$ for 80 degrees of freedom.)}
\label{pcasp}
\end{figure}

From the previous analysis, one can conclude that the source showed
little spectral and intensity variations along the entire observation
on timescales of hundreds of seconds.  We have made a PCA spectrum for
a segment of the observation using the latest versions of the
calibration matrices available at the GOF\footnote{Version 2.31 of
{\it pcarsp} was used.}. The spectrum of the source is rather hard, and
within the calibration uncertainties at low energies (below 3 keV),
can be adequately fitted with a power law of index $\sim 2.09 \pm
0.01$ ($1\sigma$ on 1 parameter, see Fig. \ref{pcasp}). The best fit
corresponds to an unabsorbed 1-20 keV flux $\sim 2 \times 10^{-9}$
\ergscm~ consistent with the flux seen by EINSTEIN during a possible
high state of the source. However, there are no doubts that the PCA
spectrum is harder than the Thermal Bremsstrahlung (TB) spectrum that
may have been observed with EINSTEIN (in particular the 6 keV TB fit
is ruled out). The PCA spectrum resembles more the power law spectra
previously observed by EXOSAT and TTM in a so-called ``low state''. A
more detailed spectral analysis including better background
subtraction (which may be an important issue for
\onee~given its location in the Galactic center region), and
combination of the HEXTE data is the subject of a separate paper
\cite{tz2:barret98aa}.

\begin{figure}[t!]
\vspace{0cm}
\hspace{-0.2cm}\centerline{\psfig{figure=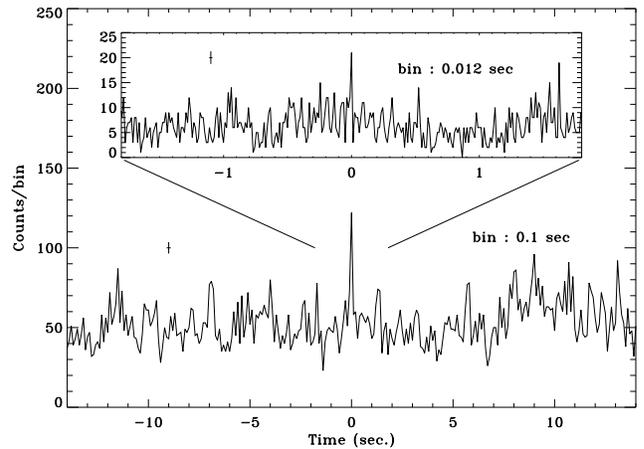,width=9.0cm}}
\vspace{0cm}
\caption{The 2-20 keV PCA light curve of 1E1724-3045 with binning 
times of 0.0122 and 0.1 second. This figure illustrates the flickering
of the source. The top panel is a zoom ($\pm 2$ sec) around the shot
shown on the main panel. The count rate is $\sim 520$ \ctss; the
background contributes to less 10\% of this value. 1$\sigma$
statistical error bar is shown for indication.}
\label{lcrapide}
\end{figure}

\section{Power Density Spectra (PDS)}

On timescales ranging from milliseconds to seconds, the source
displays intense flickering. This is illustrated in Fig \ref{lcrapide}
where we show a segment of the observation with two binning times of
0.0122 and 0.1 sec.

\begin{figure}[t!]
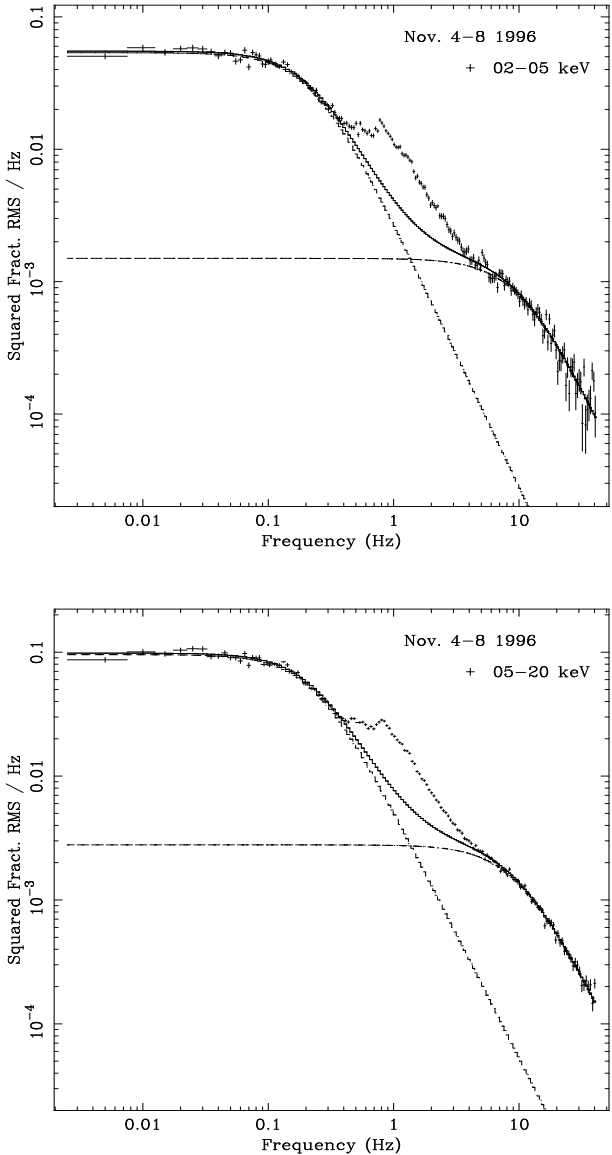
 
\centerline{\psfig{figure=fig6a.ps,height=8cm,width=9.0cm,angle=-90}}
\centerline{\psfig{figure=fig6b.ps,height=8cm,width=9.0cm,angle=-90}}
 \caption{Normalized power spectra of 1E1724-3045 in the energy band
 2-5 keV (top) and 5-20 keV (bottom). The two dashed lines represent
 the two zero-centered Lorentzian models used in the fit
 (top). Residual noise power after the model has been subtracted shows
 the broad QPO feature around 0.8 Hz. }
\label{pds2-5-5-20}
\end{figure}

In order to investigate the rapid variability of the source on such
timescales, we made PDS in the frequency range $2 \times 10^{-3} - 40$
Hz in different energy bands. Each continuous set of good data
(lasting $\sim$ 2000 sec.) was divided into $M$ segments of $N=16384$
bins of $\delta t=0.012207$ second duration. A Fast Fourier Transform
algorithm was used to convert each data segment into frequency space.
The final power spectrum of one segment was obtained by taking the
ensemble average over all $M$ segments. The power spectra were
subsequently normalized to fractional RMS densities according to
Belloni and Hasinger \cite*{belloni90aa}. For white noise substraction
we have used the expected value of 2.0 for poissonian noise. By
looking at PDS at frequencies above 700 Hz, the value observed does
not differ by more than 0.1\% of the nominal value, indicating that
deadtime corrections are negligible in our case. Afterwards, all power
spectra were rebinned logarithmically in order to increase the
statistical weight of individual frequency bins. Typical PDS spectra
are shown in figure
\ref{pds2-5-5-20} for two energy bands: 2-5 and 5-20 keV.

\subsection{Modelling of the PDS}

The shape of the PDS at low frequencies suggests that the source
displays the so-called High Frequency Noise or Low-State Noise
\cite{vanderklis95cup}. In this case the PDS can be described by a
Lorentzian centered on zero frequency, as given in the following
formula:

\begin{center}
\begin{equation}
PDS ( \nu )=\frac{K}{1+(\frac{\nu}{\nu_0})^2}
\end{equation}
\end{center}

This PDS is produced if the noise is made up of a superposition of
finite events (shots) whose characteristic duration ($\tau$) is
related to the half width half maximum ($\nu_0$) of the Lorentzian
($\tau=\frac{1}{2.\pi~\nu_0}$).  We thus decided to fit the low
frequency part of the spectrum with a Lorentzian\footnote{We are
performing the fitting of the power density spectra with XSPEC. For
that purpose, we just made fake arf and rmf files with unity diagonals
and zero elsewhere. pha files contain the PDS multiplied by the width
of the frequency bins. This has the advantage of using the efficient
interface of XSPEC for the definition of the models, the robustness of
the fitting routines, and the user friendly graphic interface.  In
addition we have found quite convenient to add our own models.}.  The
Lorentzian provides a good fit to the low frequency part of the
spectrum, in particular much better than a broken power law.  The
frequency \nuflat~above which the PDS steepens is around $\sim 0.1$
Hz.  This means that above
\nuflat$^{-1} \sim 10$ seconds there are no more time correlated
variations in the X-ray light curve of \onee.

It is clear that above $\sim 0.3$ Hz, the shape of the PDS is complex
due to the presence of a broad feature around 0.8 Hz. Regardless of
the shape of this feature, it was apparent that there was an
enhancement ``(shoulder)'' above 5-6 Hz in the PDS.  As a first
approximation, we have attempted to fit this component of the PDS with
a second Lorentzian.  It turns out that it also provides a good fit to
the data up to 40 Hz as can be seen in Figure \ref{pdsave_2_20}
(excluding the region between 0.3 and 4 Hz).  The results of the fit
are listed in Table \ref{table1}.

\begin{table}[t!]
\caption{Modeling of the Mean Power Spectrum in Various Energy Bands. 
The columns are the energy range (keV), the shot duration ($\tau_1,
\tau_2$, $\tau_3$) inferred from the first, second and third 
Lorentzians used in the fit (see text) and the integrated RMS given in
\% in the $2 \times 10^{-3}-40$ Hz. The statistical errors are quoted 
at the 1$\sigma$ level.}
\small
\label{table1}
\begin{center}
\begin{tabular}{ccccc}
\hline 
E range & $ \tau_{1} $ & $ \tau_{2} $ & $ \tau_{3} $ & $RMS_{total} $ \\ (keV) &
$msec.  $ & $ msec.  $ &  $ msec.  $ &(\%) \\
\hline
02-05 & $688 \pm 15$ & $ 15.36 \pm 0.5 $ & $\dots$ & $ 22.2 \pm 0.3 $ \\
05-10 & $694 \pm 13$ & $ 14.60 \pm 0.3 $ & $\dots$ & $ 28.1 \pm 0.3 $ \\
10-20 & $684 \pm 15$ & $ 16.22 \pm 0.6 $ & $\dots$ & $ 31.9 \pm 0.9 $ \\
20-50 & $646 \pm 40$ & $ 16.32 \pm 1.0 $ & $\dots $ & $39.9 \pm 6.0 $ \\
02-20 & $677 \pm 13$ & $ 16.50 \pm 0.1 $ & $1.4 \pm 0.2$ &$26.6 \pm 0.1 $ \\
\hline
\hline 
\end{tabular}
\end{center}
\normalsize
\end{table}

We would like to point out that other models could fit our data as
well. The Lorentzian fit as used in this work accounts for single-side
exponential shots. However, we could have also used models where the shots are
symetric (e.g. exponential rise and decay with the same timescale), in
which case the PDS should have the shape of the square of a
Lorentzian.  In this case, the PDS can be approximated at high
frequencies by a steep power law of index --4. We are not able to
distinguish between these two models because they mainly differ at
high frequencies. Unfortunately in our data, the high frequency parts
of the Lorentzians are not well resolved, due to the superposition of
these components in the PDS. Regardless of the model assumed, it is
clear that the timescales inferred for the variability will not differ
by a large factor.

We have also tried to fit the PDS with a model accounting for a
continuous time distribution of exponential shot durations $\rho (\tau
)$. If $\rho (\tau ) \propto \tau^{-2}$ between two extreme values
($\tau_1$ and $\tau_2$) then the PDS is white below $( {2 \pi
\tau_2 ) }^{-1}$, falls with a -1 slope down $( {2 \pi \tau_1 )
}^{-1}$ and with a -2 slope beyond. This model can be approximated
with the following function
\cite{shibazaki87apj}:

\small
\begin{center}
\begin{equation}
PDS ( \nu ) = \frac{K}{2\pi\nu} \left[\arctan\left(\frac{1}{2\pi\nu\tau_1}\right) - 
\arctan \left(\frac{1}{2\pi\nu\tau_2}\right)\right]
\end{equation}
\end{center}
\normalsize

\begin{figure}[t!] 
\centerline{\psfig{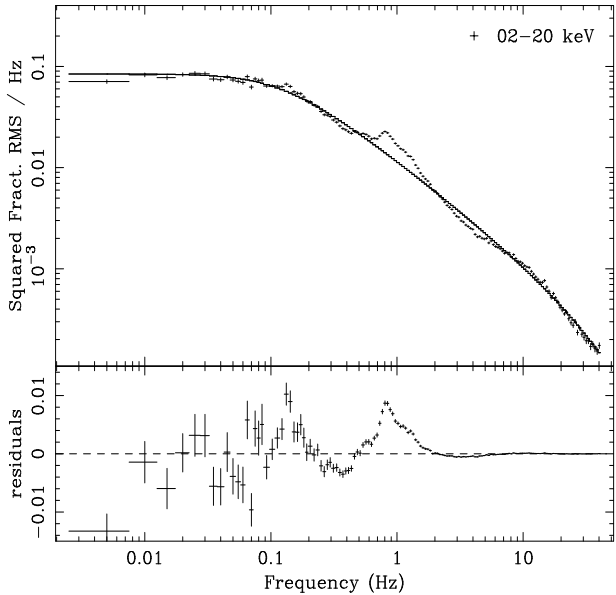}}
 \caption{Normalized power spectra of 1E1724-3045 in the energy band
 2-20 keV together with the best fit for a model accounting for
 continous distribution of shot duration (line). The bottom panel
 represents the residuals.}
\label{belhasi}
\end{figure}

This model was succesfully applied by Belloni and Hasinger
\cite*{cygx1:bh90aa} to the EXOSAT observations of Cyg X-1. Exluding 
the 0.3-4.0 Hz range, we fitted the previous model to our data for
comparison with the two lorentzian model. The fit is not as good: the
reduced $\chi^2$ is 2.3 compared to 0.7 (80 d.o.f. including a
systematic error of 5\%). As shown in Fig. \ref{belhasi}, above $\sim
2$ Hz, the fit is systematically above the data points. Furthermore
the model does not account very well for the two shoulders observed at
0.1 and above 5-6 Hz. To conclude, if the shot durations follow a
continous distribution, this function must be more complex than a
simple power law.

In the framework of the two Lorentzian model, the residuals (see
Fig. \ref{pdsave_2_20}) cannot be described by any simple models
(gaussian, etc..).  In particular the addition of a third
zero-centered Lorentzian to account for the residuals worsens the fit
as it disturbs the parameters of the Lorentzian fitting of the low
frequency part of the PDS. Clearly the way the PDS continuum is fitted
affects the shape and amplitude of the residuals. However because of
its peaked shape, by the following, we will call this feature a QPO. A
more complete modeling of the PDS, with the help of simulations and
shot profile studies will be the subject of a forthcoming paper
\cite{tz2:olive98aa}.

The main parameters of the broad QPO feature are: its frequency at
$\sim 0.8$ Hz and its total width of $\sim 3$ Hz. These two parameters
are energy independent and stable within the entire observation.

We noticed that in all PDS there is systematically a narrow and sharp
feature in the residuals around 0.15 Hz (e.g.
Fig. \ref{pdsave_2_20}). On the other hand we haven't found evidence
for the unconfirmed 0.09 Hz narrow QPO previously reported from the
EXOSAT-ME observation
\cite{belli86iau}.

\begin{figure}[t!] 
\vspace{0cm}
\hspace{0cm}\centerline{\psfig{figure=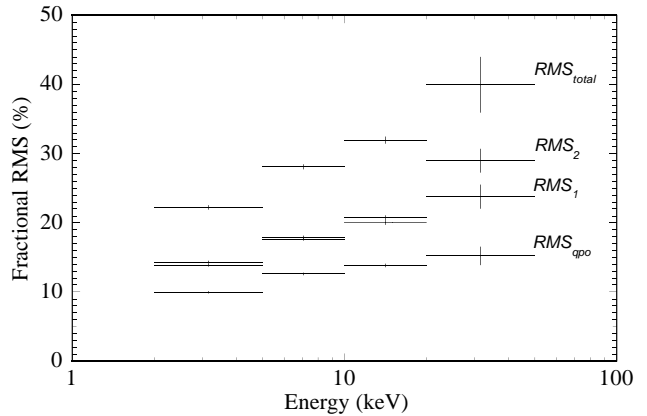,width=8.5cm}}
\vspace{0cm}
\caption{The RMS of the two shot noise components and the QPO feature. The 
total RMS is computed as the quadratic mean of the RMSs of the three
components. There is a strong positive correlation of the RMS with
energy for all three components.}
\label{rmswithenergy}
\end{figure}

The PDS made in different energy bands are all consistent in shape
with each other. For instance the values $\tau_1$ and $\tau_2$ are
constant and independent of the energy as shown in Table
1. Interestingly, the RMS of the three components shows a positive
correlation with energy as illustrated in Fig
\ref{rmswithenergy}.

\subsection{Evidence for millisecond variability in \onee ?}

In the 2-20 keV energy band which maximizes the signal to noise ratio
we made a PDS extending up to 400 Hz (i.e the binning time of the
light curve is now 1.2 milliseconds).  Above 40 Hz, in excess of the
two Lorentzian fit described above, we found a new component (see
Fig. \ref{pdsave_2_20} later). We again tried to fit this high frequency
component with a third Lorentzian.  If this model is correct, and if
interpreted in the framework of the shot noise model, the fit implies
a shot decay time of 1.4 ms.  The RMS of this third component is $6.0
\pm 0.5$\%. As can be seen the inferred shot timescale is very close
to the bin time of the light curve. We have checked through
simulations that this component is not an artefact of the binning of
light curve.

\section{Discussion}

\begin{figure}[h!] %
\vspace*{0.cm}
\hspace{0cm}\centerline{\psfig{figure=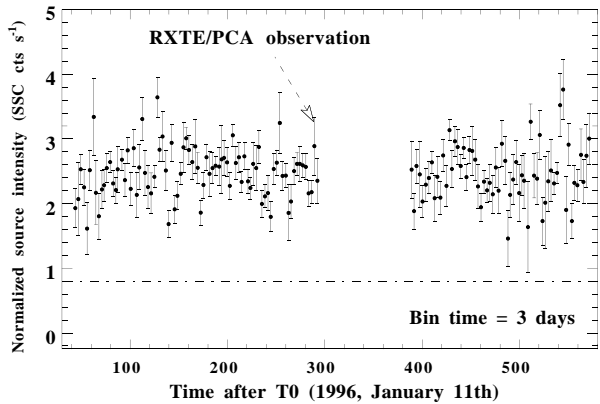,width=9.5cm}}
\vspace*{-7cm}

\caption{RXTE-ASM light curve of 1E1724-3045 showing that the source 
is steady around the level of 40 mCrab in the 2-12 keV band (2.5 SSC
\ctss). Start time is 1996, January 11th. The last point was taken on
1997, August 6th. The large gap (from MJD 50395 to 50442) is caused by
the sun traversing the Galactic Center region. During this interval,
the sun was too close to \onee~for the ASM to observe it. Our
observation took place right before the gap.}
\label{xteasm}
\end{figure}

\onee~was observed with RXTE in a hard state for which the X-ray spectrum 
is a hard power law and the PDS shows the so-called ``High Frequency
Noise''.  According to Hasinger and Van der Klis \cite*{hasinger89aa},
\onee~was observed in its ``Island state'', and remained in this
state throughout the entire observation. \onee~is a persistent X-ray
source as illustrated in Fig. \ref{xteasm} where we show its RXTE-All
Sky Monitor light curve. The mean ASM count rate corresponds to a flux
of $\sim 40$ mCrab in the 2-12 keV range fully consistent with the PCA
count rate. This figure suggests that \onee~ spends a very large
fraction of its time in the ``Island'' state with a hard X-ray power
law spectrum whose extrapolation in hard X-rays ($\ga 40$ keV) can
account for the averaged flux observed by SIGMA
\cite{tz2:barret98aa}. The low variability of the source on long
timescales (days to months) obviously agrees with the fact that
\onee~is also a persistent source of hard X-ray emission, as inferred
from SIGMA observations spanning over 6 years
\cite{gold93:2ndgro,vargas96int}.

\subsection{Origin of the shots ?}

Up to 40 Hz (and excluding the 0.3-4 Hz region), we found that the sum
of two Lorentzians was a possible fit of the PDS continuum.  In the
framework of the shot noise model, this first suggests that the light
curve of the source is made of a superposition of randomly time
distributed shots of two different timescales ($\sim 16$ msec and
$\sim 680$ msec).  Obviously, these two shot durations are too large
to be related to dynamical Keplerian timescales around the neutron
star. Comparing the competing timescales in standard Sakura-Sunyaev
accretion disks, Miyamoto et al. \cite*{miyamoto92apjl} have suggested
that shot durations less than 1 second could be related to the time
needed for the matter to flow through an X-ray emitting region of the
accretion disk where most of the energy is gravitationally
released. This viscous timescale is very sensitive to the disk model
used and the value of the viscosity parameter assumed, and might not
be appropriate to \onee~for which it is known that standard
$\alpha$-disk cannot account for the observed energy
spectra. Nevertheless, in this picture the X-ray shots would arise
from inhomogeneities approaching the inner disk region. These
inhomogeneities whose origin is unknown would then be split into rings
of matter which would cause shots of duration equal to the drift time
of the matter within the inner disk region. One prediction of that
model is that as the shot progresses and gets closer to the central
object, its spectrum should harden, at least in the case of a black
hole \cite{miyamoto92apjl}. Given the statistical quality of our data,
it might be possible to select and superpose the best time resolved
and strongest shots and follow their spectral evolution with
time. This model could also be pushed to explain the 1.4 ms shots that
we observed. On the other hand, for those shots, their duration is
also consistent with the Keplerian timescales at the inner disk
radius, or even at the neutron star surface. However, explaining the
shot duration with either timescales makes difficult to understand the
aperiodic nature of the signal.

\subsection{Timing similarities between black hole and neutron star systems}

\begin{figure}[t!]
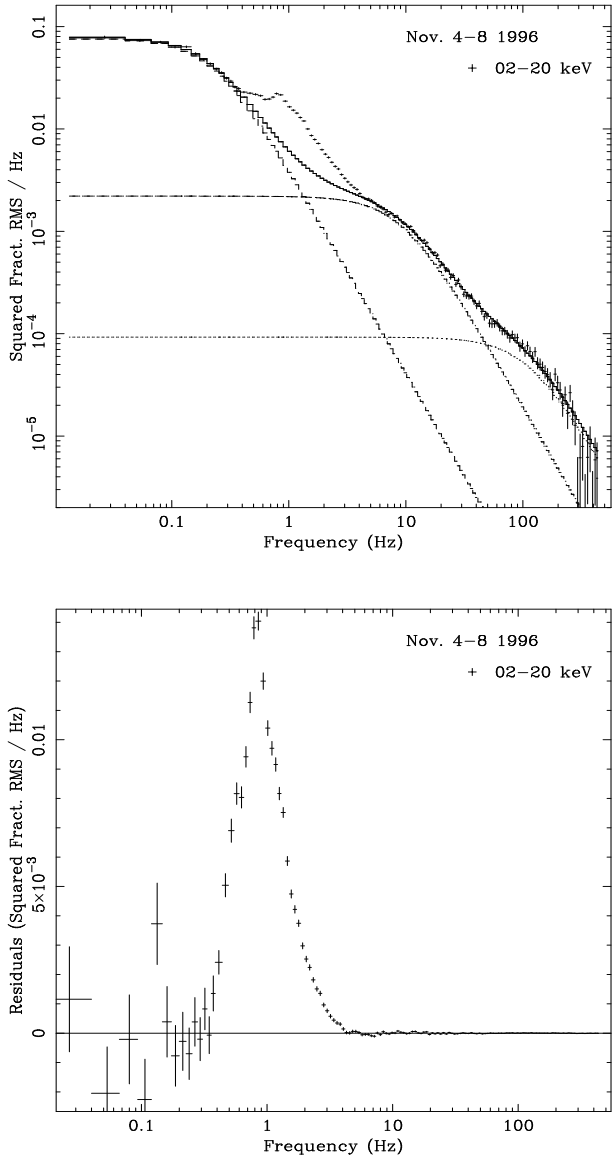
 
\centerline{\psfig{figure=fig10a.ps,height=8cm,width=9cm,angle=-90}}
\centerline{\psfig{figure=fig10b.ps,height=8cm,width=9cm,angle=-90}}
 
\caption{Time averaged normalized power density spectrum 
of the {\it neutron star system} 1E1724-3045, as observed with the PCA
in the 2-20 keV energy range. The three dashed lines represent the
three zero-centered Lorentzian models used in the fit (top). Residual
noise power after the model has been subtracted showing the broad QPO
feature around 0.8 Hz (bottom). The third Lorentzian at high frequency
(the third shoulder) provides evidence for millisecond variability in
the framework of the shot noise model.}
 
\label{pdsave_2_20}
\end{figure}

In addition, the PDS contains a broad QPO-like feature centered at 0.8
Hz. Low-frequency QPOs have been reported so far from at least two
X-ray bursters; 4U1608-522 \cite{1608:yoshida93pasj} and the somewhat
peculiar Cir X-1 \cite{cirx1:shirey96apjl}. 4U1608-522 was observed by
GINGA in a low intensity state in three observations during which the
source displayed very intense flickering. The shape of its PDS is
similar to the one shown in Fig. \ref{pdsave_2_20}.  Yoshida et
al. \cite*{1608:yoshida93pasj} modeled somewhat differently the PDS
using a sum of a broken power law, and up to three Lorentzians; one of
which accounting for the low-frequency QPO, another for a high
frequency component above $\sim 5$ Hz (similar to the second shoulder
in Fig. \ref{pdsave_2_20}). They also found a correlation between the
flat top power level and the break frequency; the so-called
``Belloni-Hasinger'' effect discovered from the black hole candidate
Cyg X-1 \cite{cygx1:bh90aa}. In addition as the source intensity
increased, the source remaining in a low state, the energy spectrum
hardened and the QPO feature as well as the high frequency component
moved to higher frequencies (from 0.4 to 2 Hz for the QPO). For Cir
X-1, the same behaviour is observed and the QPO frequency varies
within the orbital cycle from $\sim 2$ to 12 Hz
\cite{cirx1:shirey96apjl}. Unfortunately for us, \onee~was observed by
RXTE in a single spectral/intensity state, and the PDS remained very
stable in shape throughout all the observation; so we are unable to
test the presence of such a correlation in our data.

%%%%%%%%%%%%%%%%%%%%%%%%%%
\begin{figure}[t!] 
%%%%%%%%%%%%%%%%%%%%%%%%%%
\centerline{\psfig{figure=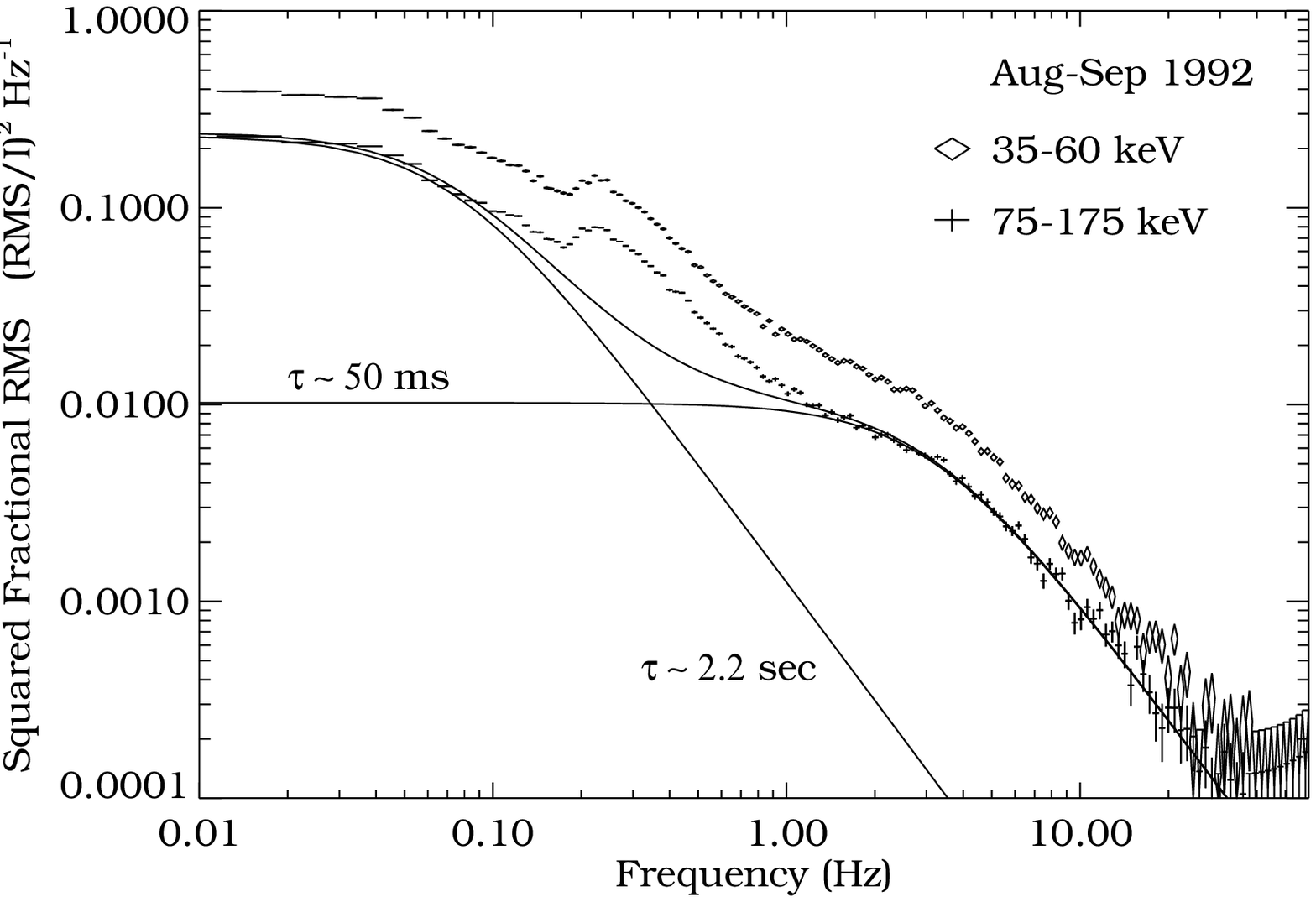,height=8cm,width=8.5cm}}
\centerline{\psfig{figure=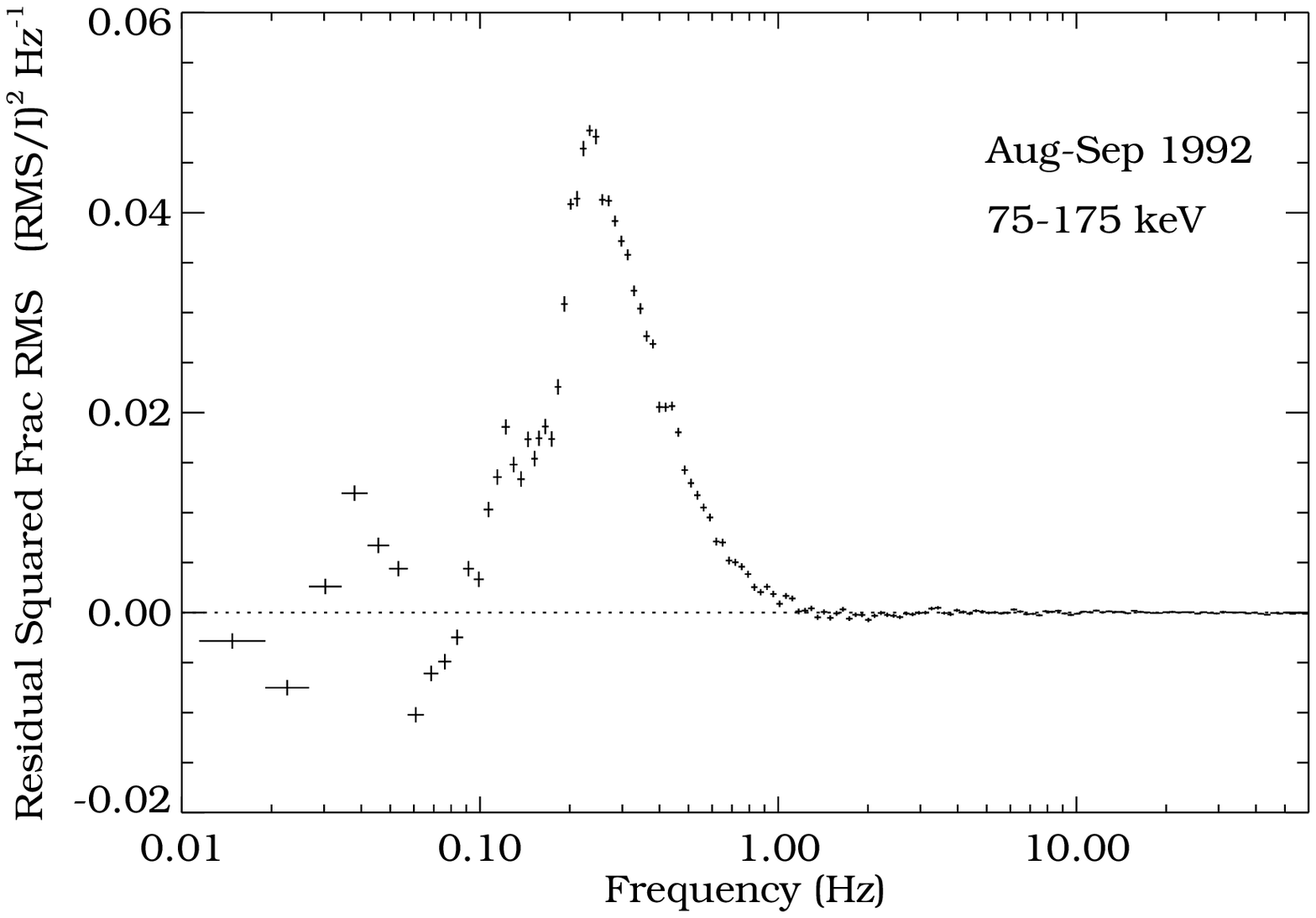,height=8cm,width=8.5cm}}
 
\caption{Normalized power spectra of the {\it black hole candidate} 
GRO J0422+32, as observed by OSSE the 35-60 and 75-175 keV bands
(top). Residual noise power in the 75-175 keV band after the model has
been subtracted (bottom) (Courtesy of Grove et al. (1993)). Note the
strong similarities with Fig. \ref{pdsave_2_20}.}
 
\label{grojosse}
\end{figure}

Atoll sources in their ``Island'' states have striking similarities
with black hole candidates both in their spectral and timing
properties. In this state, ``Atoll'' sources emit hard X-ray tails,
e.g. 4U1608-522 \cite{1608:zhang96aas}. The similarities are even more
pronounced for the timing properties.  There was already the
convincing example of 4U1608-522 and Cyg X-1
\cite{1608:yoshida93pasj,vanderklis95cup} which both show the same PDS
shape and high level of RMS ($\ga 30$\%) in their low X-ray states.
The comparison of our PCA data for \onee~with those recorded by OSSE
from GRO J0422+32 \cite{groj:grove932ndgro} provides another
spectacular example of such similarities (see
Fig. \ref{grojosse}). GRO J0422+32 is an X-ray transient which was in
outburst in 1992
\cite{groj:paciesas92iau} during which the OSSE hard X-ray (35-150 keV)
observations were performed. GRO J0422+32 is considered as a black hole
candidate with a more likely mass for the compact object in the range
2.5-5 \msol \cite{groj:casares95mnras}. Although taken at higher
energies, the PDS and the QPO feature seen in \groj~are strikingly
similar in shape to the one we observed from \onee. The only
difference between the two PDS are the shot timescales inferred from
the fitting. For GRO J0422+32 the shot timescales are respectively 50
msec and 2.2 seconds while the broad QPO feature is centered at 0.2
Hz. It is interesting to note that there is about a factor of $\sim 4$
for $\tau_1, \tau_2$, \nuqpo$^{-1}$ between \onee~and \groj.

Low frequency QPOs have been observed from several other black hole
candidates; of the most convincing detections we find GX339-4 (0.77
Hz) \cite{gx339:grebenev91sal}, LMC X-1 (0.08 Hz)
\cite{lmcx1:ebisawa89pasj}, Cyg X-1 (0.04 Hz)
\cite{cygx1:angelini92iau}. Note that for the few
cases we have so far, it seems that QPO frequencies from neutron star
systems ($\ga 0.5$ Hz) seem on average to be larger than those seen in
black hole systems.  It is unlikely though that the QPO frequency is
directly correlated to the mass of the compact object (the larger the
mass, the slower the QPOs), as in this case, we would expect it to be
comparable for all neutron star systems, and constant within a given
system. As said above, this is not the case neither for 4U1608-522 nor
for Cir X-1 \cite{1608:yoshida93pasj,cirx1:shirey96apjl}.

\subsection{Possible common origin for the low frequency QPOs in black holes 
and neutron stars?}
 
The HFN with a high level of RMS on to which a low frequency QPO is
superimposed thus emerges as a similarity between black hole and
neutron star systems, suggesting that similar mechanisms are involved
independently of the nature of the compact object. The origin of these
QPOs is still unclear. It is clear however that they cannot be related
to Keplerian motions as the associated radius ($\sim 6000$ km) is much
larger than the radius of the region where the X-rays are emitted
($\la$ 300 km). Given the short shot durations inferred from our
analysis, it appears unlikely that the QPO can result from the direct
modulation of the shots. In addition the large RMS, and its positive
correlation with energies implies that it originates from the region
of main energy release (i.e close to the compact object).  It is
rather difficult to find a mechanism which could cause QPOs in that
region. Depending on the viscosity parameters, disk luminosity
oscillations around 1 Hz might result from thermal-viscous
instabilities developing within the inner parts of the disk around a
neutron star \cite{chentaam94apj}. These oscillations will exhibit
quasi-periodic behaviour if the mass flow entering the inner disk
region is not constant.

Independently of the mechanism producing the shots Vikhlinin et al.
\cite*{vikhlinin94aa} have suggested that the low frequency QPOs could
result from a weak interaction between the shots when the instability
develops in a region of stable energy supply. To keep the energy
released constant on large timescales, the appearance of strong shots
should affect the amplitude and/or the probability of occurrence of
subsequent shots. In our case, this would mean that on average, one
shot is followed about $\sim 1.25$ seconds later by a second shot.

\subsection{Low frequency QPOs and the inner disk precession ?}

Beside the two previous models which do not make any assumptions on
the nature of the compact object, there is another one which involves
a rapidly rotating weakly magnetized neutron star. This model has
recently been proposed by Stella and Vietri \cite*{stella97apjl}. The
latter authors showed that the broad peaks (around tens of Hz)
observed in the PDS of several Atoll sources (4U0614+091, MXB1728-34,
and KS1731-260), all displaying twin kilo-Hz QPOs, could be due to
the precession of the innermost disk region.  So close to the neutron
star, there are two precession frequencies to consider; one caused by
the ``frame dragging'' effect \cite{lense} (\nult), and another one
induced by the oblatness of a fastly rotating neutron star
\cite{kahn} (\nucl). 

\begin{figure}[t!] 
\vspace*{-2cm}
\centerline{\psfig{figure=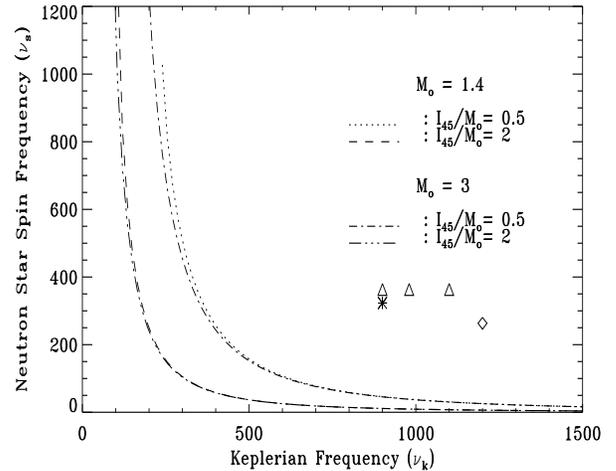,height=9cm,width=9.50cm}}
\caption{Allowed range of \nus~and \nuk~which would match a 0.8 Hz QPO 
for different values of \iquacin/\mzero, and \mzero~(see text).
According to the formula in Stella and Vietri (1997), we assign $\cos
\beta$ to its mean value between 0 and $\frac{\pi}{2}$ ($\beta$ tilt
angle of the equatorial plane in \nucl, and a dimensionless parameter
$\eta = 10^{-2}$). Also indicated \nus~and \nuk~observed from
4U0614+091 (star), KS1731-260 (square) and MXB1728-34 (triangles).}
\label{precess}
\end{figure}

The precession frequency is the sum of the two
frequencies, in which \nult~generally dominates over \nucl.  Both
frequencies depend on the neutron star spin frequency (\nus) and the
Keplerian frequency at the inner disk radius (\nuk).  The latter
quantities can be estimated if the observed twin kilo-Hz QPOs are
interpreted with a magnetospheric beat frequency model
\cite{alpar85nat}. The neutron star parameters (mass,
M=\mzero~\msol~and moment of inertia, I=$10^{45}$ \iquacin) are two
other, and more uncertain parameters which critically determine the
range of precession frequencies.  For neutron star models with stiff
equation of states, \iquacin/\mzero=2 and for \mzero $\sim 2$
\cite{cook94apj,friedman86apj}, the precession model recovers
successfully the frequency of the peaks observed in all three sources
(between 10 and 40 Hz) \cite{stella97apjl}. It is thus tempting to
apply this model to our data, although the QPO from \onee~is at a
significantly lower frequency. Furthermore, in our case, we do not
have yet a value for \nus and \nuk\footnote{Although not yet complete,
our analysis has failed so far to reveal any significant high
frequency QPOs above 300 Hz. The results of the complete analysis will
appear elsewhere.}. Yet, if one assumes \mzero~and \iquacin, in the
framework of the previous model, one can estimate the range of
\nus~and \nuk~which would match a QPO frequency of 0.8 Hz. This is
shown in Fig. \ref{precess} for which we have computed \nus~and
\nuk~for plausible ranges of \iquacin~such that $0.5 \le$
\iquacin $\le 2.0$ and \mzero~(1.4 $\le $ \mzero $\le
3.0$) \cite{cook94apj,friedman86apj}. This figure illustrates well the
weak dependence of \nus~and \nuk~ against the neutron star mass, and
the stronger dependence on \iquacin/\mzero. However, none of the
curves intersects with the region where \nus~and \nuk~have been
observed for the three sources listed above (900 $\la$ \nuk $\la
1200$, and 300 $\la$ \nus $\la 500$ Hz). This might have something to
do with the fact that we don't see any obvious high frequency QPOs
from this source.

\section{Conclusions} 

Our RXTE/PCA observation enabled us to classify \onee~as an ``Atoll''
source.  It was observed in its ``Island'' state for which its X-ray
energy spectrum is hard and the PDS displays high frequency noise with
a high level of RMS ($\sim$ 30\%). The shape of the PDS is complex but
can be decomposed in the framework of the shot noise model.  In
addition the PDS contains a QPO-like feature peaking at $\sim 0.8$ Hz.
Similar PDS properties have already been observed from several black
hole candidates, and in particular from \groj~for which the
similarities are striking.  A complete timing analysis of the PCA data
is in progress with the aim of studying the shot profiles, the origin
of the QPO, the spectral evolution along the shots, time lags as a
function of energy, etc... \cite{tz2:olive98aa}.  This illustrates
well the wealth of the PCA data.  Some very valuable information will
also be provided by the HEXTE experiment which will for the first time
enable us to study the hard power law tail and the aperiodic
variability simultaneously from X-rays to hard X-rays.

% Undoubtedly,
%the data will help us in understanding the complex accretion process
%onto this neutron star system.

\begin{acknowledgements}
We wish to thank Drs Keith Jahoda and Charles Day of the RXTE Guest
Observer Facility for their valuable help in the course of the
analysis. We are very grateful to J.E. Grove for providing us with
Figures \ref{grojosse}. This research has made use of data obtained
through the High Energy Astrophysics Science Archive Research Center
Online Service, provided by the NASA-Goddard Space Flight Center. The
RXTE/ASM light curve of \onee~was downloaded from the Web site
http://space.mit.edu/XTE/ASM-lc.html at Massachussets Institute of
Technology. The authors are grateful to S.N. Zhang for helpful
discussions. \\ We also thank the referee, Dr. G. Hasinger, for his
careful and helpful review of the paper. 
\end{acknowledgements}

%\bibliography{/home/barret/Latex/name_journals,tz2,/home/barret/Latex/db,/home/barret/Latex/xraybursters}
%\bibliographystyle{/home/barret/Latex/ASTRON/astron2} 

\end{document}